# High-Performance $KV_3Sb_5/WSe_2$ van der Waals Photodetectors

Yang Yang [1,2], Shaofeng Rao [1,2], Yuxuan Hou [1,2], Jiabo Liu [1,2], Deng Hu [3,4], Yufei Guo [1,2], Jianzhou Zhao [1,2], Hechen Ren [1,2], Zhiwei Wang [3,4 †], and Fan Yang [1,2*]

[1] Center for Joint Quantum Studies and Department of Physics, School of Science, Tianjin University, Tianjin 300350, China

[2] Tianjin Key Laboratory of Low Dimensional Materials Physics and Preparing Technology, Department of Physics, Tianjin University, Tianjin 300350, China

[3] Key Laboratory of Advanced Optoelectronic Quantum Architecture and Measurement, Ministry of Education, School of Physics, Beijing Institute of Technology, Beijing 100081, China

[4] Micronano Center, Beijing Key Lab of Nanophotonics and Ultrafine Optoelectronic Systems, Beijing Institute of Technology, Beijing 100081, China

[†] Corresponding author: zhiweiwang@bit.edu.cn

[*] Corresponding author: fanyangphys@tju.edu.cn

**Keywords**: kagome metal, $KV_3Sb_5$, van der Waals heterojunction, photodetector

## Abstract:

Kagome metals $AV_3Sb_5$ (A = K, Rb, Cs) have recently emerged as a promising platform for exploring correlated and topological quantum states, yet their potential for optoelectronic applications remains largely unexplored. Here, we report high-performance photodetectors based on van der Waals $KV_3Sb_5/WSe_2$ heterojunctions. A high-quality Schottky interface readily forms between $KV_3Sb_5$ and $WSe_2$, enabling efficient separation and transport of photoinduced carriers. Under 520 nm illumination, the device achieves an open-circuit voltage up to 0.6 V, a responsivity of 809 mA $W^{-1}$, and a fast response time of 18.3 μs. This work demonstrates the promising optoelectronic applications of Kagome metals and highlights the potential of $KV_3Sb_5$-based van der Waals heterostructures for high-performance photodetection.

## 1. Introduction

Recently, the layered kagome metals $AV_3Sb_5$ (A = K, Rb, Cs)[1-8] have drawn significant research attention owing to their unique lattice geometry and unconventional electronic properties. These $AV_3Sb_5$ compounds crystallize in a layered structure, within which the vanadium atoms form a two-dimensional (2D) kagome network[1,7]. Such a distinctive atomical configuration directly leads to a range of intriguing electronic features, including flat bands, Dirac dispersions, and topological band structures[3,5]. As a prototype member of the $AV_3Sb_5$ family, $KV_3Sb_5$ hosts a variety of emergent electronic orders, including superconductivity with a bulk $T_c$ of 0.93 K[2,8], an unconventional charge-density-wave (CDW) phase with possible chiral characteristics[9-12], as well as nontrivial topological states originating from the kagome lattice[2,13]. These features make $KV_3Sb_5$ an ideal platform for investigating the interplay among superconductivity, CDW order and band topology.

Despite these advances, research on $KV_3Sb_5$ has so far focused mostly on its intrinsic quantum phases, leaving its potential for optoelectronic applications largely unexplored. Benefiting from its relatively high carrier mobility, atomically flat surfaces, and semi-metallic band structure, $KV_3Sb_5$ is also expected to serve as a promising charge-transport layer in van der Waals heterostructure devices. Nevertheless, no photodetector or other optoelectronic device based on $KV_3Sb_5$ has been demonstrated to date.

In this work, we report the first photodetector based on a van der Waals $KV_3Sb_5/WSe_2$ heterojunction. A high-quality Schottky interface naturally forms between $KV_3Sb_5$ and $WSe_2$, enabling efficient separation and extraction of photogenerated carriers. Under 520 nm illumination, the device achieves a high open-circuit voltage up to 0.6 V and a responsivity of 800 mA $W^{-1}$, together with fast photoresponse dynamics characterized by a rise time of 18.3 μs and a decay time of 7.9 μs. Notably, these performance metrics surpass those of most previously reported $WSe_2$-based van der Waals photodetectors[14-29]. Our results not only establish $KV_3Sb_5$ as an effective charge-transport component in hybrid optoelectronic architectures but also open new opportunities for integrating kagome metals into functional optoelectronic devices.

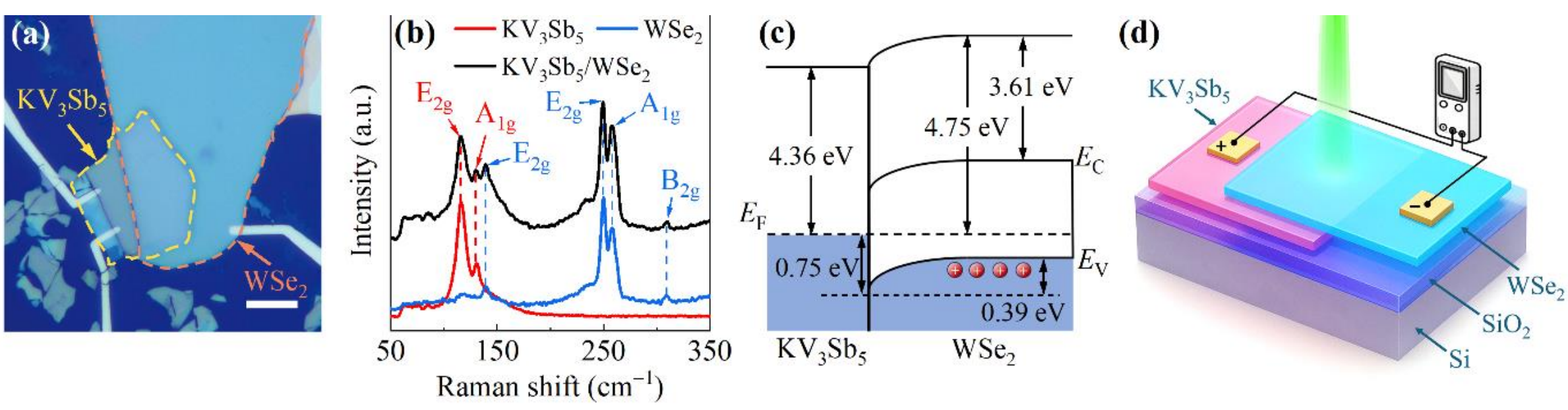

**Figure 1**. a) Optical microscope image of a typical $KV_3Sb_5/WSe_2$ photodector (scale bar: 20 μm; junction area: 1094 $\mu m^2$). b) Raman spectra of pristine $KV_3Sb_5$ and $WSe_2$ flakes, as well as the stacked $KV_3Sb_5/WSe_2$ bilayer, with the characteristic vibrational modes of each material indicated. c) Schematic band diagram of the $KV_3Sb_5/WSe_2$ heterostructure. d) Schematic device structure and measurement configuration.

## 2. Results and Discussion

### 2.1 Device structure

The photodetectors investigated in this work are based on vertically stacked $KV_3Sb_5/WSe_2$ van der Waals heterostructures. To ensure high-quality interfaces, $WSe_2$ flakes were exfoliated and transferred onto freshly cleaved $KV_3Sb_5$ flakes inside an Ar-filled glove box, yielding clean and well-defined interfaces in the overlapping region. Pt electrodes were fabricated on both the $KV_3Sb_5$ and $WSe_2$ layers to provide reliable electrical contacts for optoelectronic measurements. Details of the device fabrication procedure are described in the Methods section. Figure 1a shows an optical microscope image of a representative device. Unless otherwise specified, all data presented in the main text are obtained from this device. Additional results from an independent device are presented in the Supporting Information. This device exhibits photoresponse characteristics fully consistent with those discussed in the main text, demonstrating good data reproducibility.

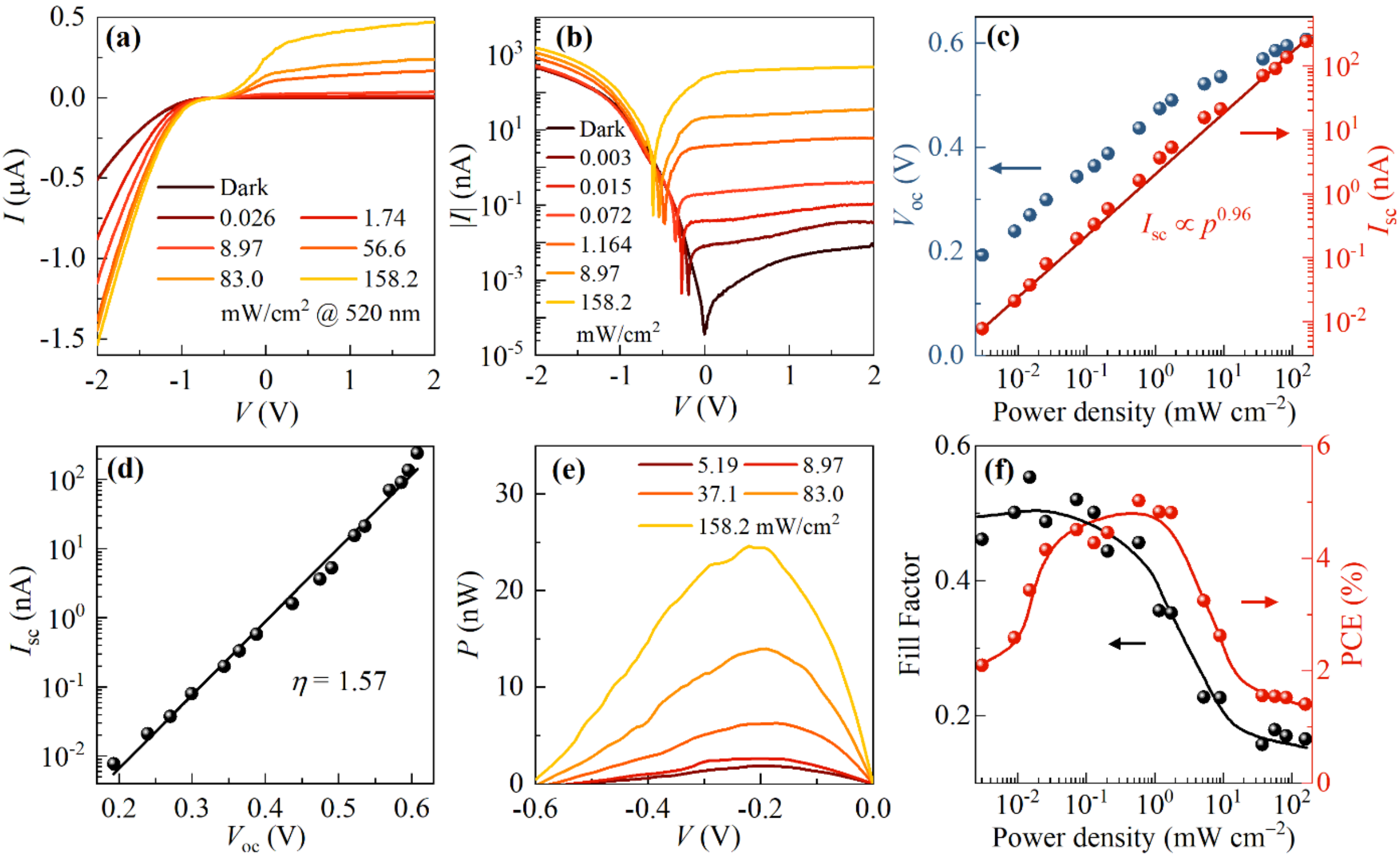

**Figure 2**. a) Linear- and b) logarithmic-scale $I$-$V$ curves measured under 520 nm illumination at various illumination power densities. c) Open-circuit voltage ($V_{oc}$) and short-circuit current ($I_{sc}$) plotted as functions of illumination intensity. d) $V_{oc}$-$I_{sc}$ correlation, where a fit to the data yields a diode ideality factor of $\eta$ = 1.57. e) Output power as a function of bias voltage. f) Fill factor and photoelectric conversion efficiency (PCE) as functions of illumination intensity.

The high quality of the $KV_3Sb_5/WSe_2$ interface is further confirmed by Raman spectroscopy. Figure 1b shows the Raman spectra of a $KV_3Sb_5/WSe_2$ stack, which exhibits clear characteristic Raman peaks at 116 and 130.3 $cm^{-1}$ in the $KV_3Sb_5$ region and peaks at 139.5, 249.7, 258.2, and 310 $cm^{-1}$ in the $WSe_2$ region, in good agreement with previous reports[30-32]. In the overlapping region, all characteristic Raman peaks of $KV_3Sb_5$ and $WSe_2$ are clearly visible without discernible shifts, indicating the formation of a clean van der Waals interface without strain or interlayer coupling strong enough to modify the vibrational signatures[33,34].

Due to the weak van der Waals interaction between the K and Sb layers, the exfoliated $KV_3Sb_5$ surfaces can be either K- or Sb-terminated[35-38]. Our first-principles calculations reveal that both terminations yield an identical work function of 4.36 eV for $KV_3Sb_5$. In contrast, $WSe_2$ is a p-type semiconductor with a higher work function of about 4.75 eV[39]. Such a difference in work function naturally leads to the formation of a Schottky junction at the $KV_3Sb_5/WSe_2$ interface, as schematically illustrated in Figure 1c. Upon illumination, photogenerated carriers are efficiently separated by the built-in electric field at the Schottky interface, giving rise to a pronounced photovoltaic effect. Figure 1d presents a schematic of the device architecture and measurement configuration. During measurements, the device is uniformly illuminated by a 520 nm laser, and the electrodes on $KV_3Sb_5$ and $WSe_2$ are respectively connected to the positive and negative terminals of a DC sourcemeter or oscilloscope for electrical measurements.

## 2.2 Current-Voltage Characteristics under Illumination

The current–voltage ($I$–$V$) characteristics of the $KV_3Sb_5$/$WSe_2$ photodetector were investigated under 520 nm illumination. Figures 2a and 2b present the $I$–$V$ curves measured in the dark and under various illumination power densities. In the dark, the device exhibits pronounced rectifying behavior, with significant current conduction only under negative bias, indicating the formation of a Schottky junction between semiconducting $WSe_2$ and semimetallic $KV_3Sb_5$. The rectification ratio reaches approximately $6 \times 10^4$ at $\pm 2$ V, reflecting the presence of a strong built-in interfacial electric field, which is essential for photovoltaic operation.

Upon illumination, , the device shows a strong photovoltaic response at zero bias. At an incident power density of 158.2 mW cm$^{-2}$, the open-circuit voltage and short-circuit current reach $V_{\mathrm{oc}} = 0.6$ V and $I_{\mathrm{sc}} = 242$ nA, respectively, as shown in Figure 2c. With increasing illumination intensity, $V_{\mathrm{oc}}$ increases logarithmically and saturates near 0.6 V, while $I_{\mathrm{sc}}$ scales nearly linearly with the incident power ($I_{\mathrm{sc}} \propto p^{0.96}$). This near-unity exponent indicates few trap states and efficient separation and extraction of photogenerated carriers at the Schottky interface[40].

In addition to the photovoltaic response at zero bias, the $KV_3Sb_5$/$WSe_2$ device also exhibits a clear photoconductive response under finite bias voltages. As shown in Figure 2b, illumination leads to a pronounced increase in current over the entire measured bias range. At an applied bias of $V = 2$ V, the on/off current ratio exceeds $10^4$, indicating efficient photoinduced modulation of the device conductance.

To further analyze the transport property of the $KV_3Sb_5$/$WSe_2$ Schottky junction, we plot the $I_{\mathrm{sc}}$-$V_{\mathrm{oc}}$ data in Figure 2d and performed curve fitting using the standard diode equation[41]

$$V_{\mathrm{oc}} = \frac{\eta k_{\mathrm{B}} T}{e} \ln\left(\frac{I_{\mathrm{sc}}}{I_{\mathrm{sr}}} + 1\right), \tag{1}$$

where $\eta$ is the ideality factor, $I_{\mathrm{sr}}$ is the saturation current, $T$ is the temperature, $k_{\mathrm{B}}$ is the Boltzmann constant, and $e$ is the elementary charge. From the fitting, an ideality factor of $\eta = 1.57$ is extracted, which falls within the range commonly reported for high-quality van der Waals heterostructures[23,24,29,39,42]. Although an ideality factor larger than unity indicates deviations from the ideal thermionic-emission limit, such a moderate value suggests that carrier transport across the junction is still predominantly governed by thermionic emission, with only minor contributions from interface recombination or tunneling processes, consistent with a relatively clean and well-defined interface.

Overall, the $KV_3Sb_5$/$WSe_2$ device exhibits robust optoelectronic performance, enabling possible applications in both photovoltaic energy-harvesting and photodetection regimes. In the following sections, we therefore analyze the device performance separately in terms of power conversion and photodetection to further clarify its functionality in these two regimes.

## 2.3 Power Conversion Capability

To evaluate the power conversion capability of the device, we analyzed the electrical output power $P_{\mathrm{el}} = IV$ under various illumination intensities, as plotted in Figure 2e. In addition, the fill factor (FF) and power conversion efficiency (PCE) were extracted from the measured $I$–$V$ curves, as summarized in Figure 2f. Here, the FF and PCE are defined as $\mathrm{FF} = P_{\mathrm{max}}/(I_{\mathrm{sc}}V_{\mathrm{oc}})$[43] and $\mathrm{PCE} = P_{\mathrm{max}}/P_{\mathrm{in}}$[44], where $P_{\mathrm{max}}$ is the maximum output power and $P_{\mathrm{in}}$ is the incident optical power. These two parameters are conventionally used to assess the performance of photovoltaic solar cells; however, here they mainly serve as quantitative indicators of the junction quality and bias-dependent power-generation capability.

As shown in Figure 2f, both FF and PCE exhibit a pronounced dependence on illumination intensity. At low power densities below 1 mW $cm^{-2}$, the device maintains a relatively high FF of approximately 0.5, reflecting efficient carrier separation and collection at the Schottky interface. With increasing illumination intensity, FF gradually decreases, reaching about 0.16 at 100 mW $cm^{-2}$. Correspondingly, PCE increases at low illumination and reaches a maximum of ~5% around 1 mW $cm^{-2}$, followed by a decrease to ~1.7% at higher illumination levels. The decrease of FF and PCE at elevated illumination intensities can be mainly attributed to series-resistance effects originating from the $WSe_2$ channel and Pt contacts[45,46]. Additional contributions from interfacial recombination and photoinduced heating may also play a secondary role.

Although the power conversion efficiency of the $KV_3Sb_5/WSe_2$ device is limited at high illumination intensities, its pronounced photovoltaic response and high open-circuit voltage make the device particularly suitable for self-powered photodetection. Moreover, under finite bias, the device exhibits a clear photoconductive response, which enables multiple detection modes within a single device architecture. In the following section, we therefore focus on the photodetection performance of the $KV_3Sb_5/WSe_2$ device, including its responsivity and temporal response, to assess its potential for practical photodetector applications.

## 2.4 Photoresponse Characteristics

In this section, we focus on the optoelectronic detection response of the $KV_3Sb_5/WSe_2$ device. Based on the photovoltaic characteristics discussed above, we now investigate how the $KV_3Sb_5/WSe_2$ device operates as a photodetector. Three key aspects relevant to photodetection performance are considered: self-powered operation at zero bias, bias-enhanced photoconductive response, and temporal dynamics of photocurrent. Together, these results provide a comprehensive picture of the device performance under different illumination and bias conditions.

Figure 3a presents the spatial photocurrent mapping of the device under 405 nm illumination. A photocurrent is observed exclusively in the overlapping region between $KV_3Sb_5$ and $WSe_2$ whereas negligible response is detected in the individual constituent regions. This spatially localized photoresponse indicates that photocurrent generation primarily originates from the $KV_3Sb_5/WSe_2$ heterojunction rather than from the isolated materials. Such behavior is consistent with the formation of an effective Schottky junction at the interface, highlighting the role of the interfacial band alignment in facilitating photo-carrier separation and collection.

Having established the interfacial origin of the photoresponse, we next examine the photovoltaic performance of the device at zero bias. As shown in Figure 3b, the $KV_3Sb_5/WSe_2$ device exhibits stable and repeatable on/off switching under periodic illumination at zero bias, demonstrating reliable self-powered photodetection. The responsivity, defined as $R_I = I_{\rm ph}/P_{\rm in}$ [47], reaches 236 and 155 mA $W^{-1}$ under low and high illumination intensities, respectively, confirming its robust photovoltaic performance.

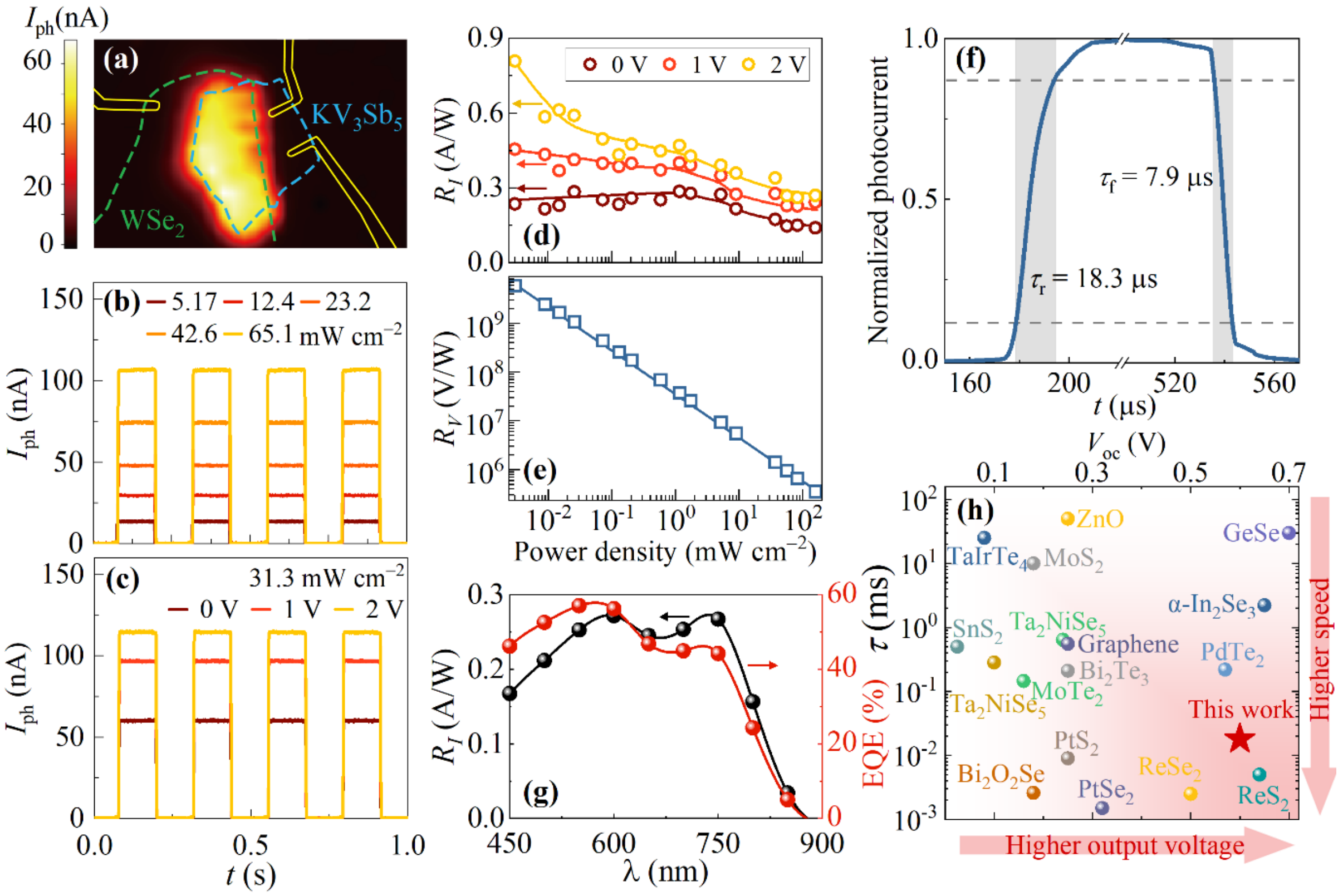

**Figure 3.** a) Photocurrent mapping under 405 nm illumination. b) Photocurrent switching characteristics at $V = 0$, measured under various illumination intensities. c) Photocurrent switching characteristics under 31.3 mW $cm^{-2}$ illumination, measured at various bias voltages. d) Current responsivity ($R_I$) under various bias voltages, and e) voltage responsivity ($R_V$) as functions of illumination power density. f) Temporal photoresponse under 83 mW $cm^{-2}$ illumination, with the rise and decay times indicated. g) Wavelength dependence of zero-bias $R_I$ and external quantum efficiency (EQE). h) Comparison of the open-circuit voltage ($V_{\rm oc}$) and photoresponse time ($\tau$) of the $KV_3Sb_5/WSe_2$ heterojunction with other previously reported X/$WSe_2$ van der Waals photodetectors, where X represents the materials indicated in the figure.

Beyond self-powered operation, the device also maintains stable photocurrent switching under forward bias, as illustrated in Figure 3c. The application of a forward bias voltage significantly enhances the photocurrent, leading to an overall stronger photoconductive response. As summarized in Figure 3d, at low illumination intensities, the current responsivity increases from 236 mA $W^{-1}$ at zero bias to 809 mA $W^{-1}$ under a 2 V bias, reflecting a substantial photoconductive gain relative to the self-powered regime.

In addition to the current-based response, the $KV_3Sb_5/WSe_2$ device also exhibits a high voltage responsivity in the self-powered mode. As shown in Figure 3e, the voltage responsivity, defined as $R_V = V_{\rm oc}/P_{\rm in}$, reaches up to $5.8 \times 10^9$ V $W^{-1}$ under low illumination conditions. This large voltage responsivity further highlights the potential of the $KV_3Sb_5/WSe_2$ device for self-powered photodetection applications.

Another key advantage of the $KV_3Sb_5/WSe_2$ photodetector lies in its ultrafast temporal response. As shown in Figure 3f, high-resolution transient measurements reveal a rise time of 18.3 μs and a decay time of 7.9 μs at zero bias, indicative of rapid photocarrier generation and extraction. Importantly, this fast response is largely preserved under applied bias (see Supporting Information), demonstrating that the device maintains rapid photoresponse in both photovoltaic and photoconductive modes of operation.

To determine the effective operating spectral range, the spectral photoresponse of the $KV_3Sb_5/WSe_2$ device was investigated using a supercontinuum light source. As summarized in Figure 3g, the device exhibits consistently high responsivity and external quantum efficiency (EQE) over the wavelength range from 450 to 800 nm. In particular, the EQE remains above 44% for wavelengths below 750 nm, reflecting efficient photon-to-electron conversion throughout the visible spectrum. At longer wavelengths, the photoresponse decreases markedly; at 850 nm, the zero-bias responsivity drops to 34 mA $W^{-1}$ and the EQE to 4.9%, indicating that the $KV_3Sb_5/WSe_2$ device is primarily optimized for visible-light photodetection.

In Figure 3h, we compare the overall performance of the $KV_3Sb_5/WSe_2$ device with that of other previously reported $WSe_2$-based van der Waals photodetectors in terms of saturated open-circuit voltage and photoresponse time. Although these two metrics originate from different application contexts, their combined comparison provides a good assessment of the device's ability to simultaneously deliver strong photovoltaic output and fast temporal response. This analysis places the present device in a favorable position within the existing performance landscape. Except for the $ReS_2/WSe_2$ heterojunction[48], the $KV_3Sb_5/WSe_2$ device outperforms the reported counterparts in at least one of these two key metrics, exhibiting either a larger open-circuit voltage or a faster response time.

For self-powered photodetectors, simultaneously achieving a high open-circuit voltage and a fast temporal response is a challenging task, as these two metrics are often constrained by competing device physics. In this context, the balanced performance of the $KV_3Sb_5/WSe_2$ heterojunction highlights the advantages of the kagome-metal/semiconductor interface, where an efficient built-in field formation enables strong photovoltaic response without sacrificing carrier extraction dynamics. These results demonstrate the potential of kagome-metal/semiconductor heterostructures as a promising platform for high-performance, self-powered photodetection.

## 3. Conclusion

In summary, we have developed a high-performance photodetector based on a van der Waals heterojunction established between the kagome metal $KV_3Sb_5$ and the semiconductor $WSe_2$. A naturally formed, high-quality Schottky interface enables efficient separation and collection of photogenerated carriers, yielding excellent optoelectronic performance under 520 nm illumination. The device exhibits a notable open-circuit voltage of 0.6 V, a high responsivity of 809 mA $W^{-1}$, and rapid photoresponse dynamics (rise time: 18.3 μs; decay time: 7.9 μs), surpassing most previously reported $WSe_2$-based photodetectors. These results validate $KV_3Sb_5$ as an effective charge-transport component in van der Waals optoelectronic architectures and bridge the gap between the fascinating physics of kagome metals and their device applications. Our findings open new opportunities for $AV_3Sb_5$-based heterostructures in high-performance optoelectronic devices.

## 4. Methods

*Crystal Growth*

The $KV_3Sb_5$ single crystals were synthesized using a self-flux method. High-purity K bulk (99.95%), V flakes (99.9%), and Sb granules (99.9999%) were weighed in a molar ratio of K: V: Sb = 7.7: 3: 14 and loaded into an alumina crucible. The V and Sb starting materials were pretreated by high-temperature hydrogen reduction to remove surface oxides. Owing to the high saturated vapor pressure and corrosiveness of K at elevated temperatures, a double-quartz-tube sealing technique was employed. The sealed crucible was then placed in a box furnace, heated to 1000 °C over 100 hours, held for 24 hours, and cooled to 400 °C over 300 hours. After cooling to room temperature, the product was retrieved and immersed in deionized water to dissolve the flux. Finally, high-quality $KV_3Sb_5$ single crystals with hexagonal shape were obtained. The $WSe_2$ single crystals used in this experiment were commercially purchased from HQ Graphene (The Netherlands).

*Device Fabrication*

Si (100) substrates with a 300 nm $SiO_2$ layer were cleaned by sequential ultrasonication in acetone and isopropanol for 10 minutes each, followed by a 120-s $O_2$-plasma treatment (200 W, 29 mTorr) to remove possible organic contaminants. $KV_3Sb_5$ flakes were exfoliated from bulk crystals using low-residual PVC protective tape and transferred onto the pre-cleaned $Si/SiO_2$ substrates. Thin $WSe_2$ flakes were first exfoliated from bulk samples using heat-release PDMS stamps and then aligned and stacked onto the selected $KV_3Sb_5$ flakes using a homemade transfer system. After the transfer, Pt electrodes were patterned by UV photolithography, followed by Pt sputtering (3 mTorr Ar, 25 W) and a standard lift-off process. To avoid oxidation, the exfoliation, transfer, and lift-off procedures were all performed inside an Ar glove box ($O_2$ and $H_2O$ < 0.01 ppm). Finally, the completed device was spin-coated with S1813 photoresist, serving as a protective layer against oxidation during the measurements.

*Device Characterization and Photoelectrical Measurement*

The Raman spectra of the $KV_3Sb_5/WSe_2$ bilayer were measured with a Horiba LabRAM HR Evolution Raman microscope under 532-nm laser excitation. Current–voltage characteristics were measured using a Keithley 2636B source meter, and the temporal photoresponse was captured with an Agilent 54832B digital oscilloscope (1 GHz bandwidth). Wavelength-dependent photocurrent measurements were performed under continuous-wave illumination from a Zolix Gloria-SC-N supercontinuum light source.

*Work Function Calculation*

The work functions of K- and Sb- terminated $KV_3Sb_5$ were calculated using density functional theory (DFT) implemented in the Vienna *ab initio* Simulation Package (VASP)[49]. The projector augmented-wave (PAW) method and the Perdew–Burke–Ernzerhof (PBE) generalized gradient approximation were employed, with van der Waals corrections incorporated via the DFT-D3 method. A plane-wave energy cutoff of 400 eV was adopted, and the total energy and Hellmann–Feynman forces were converged to $10^{-6}$ eV and 0.01 eV $Å^{-1}$, respectively. To model the $KV_3Sb_5$ surfaces, slabs were constructed with a vacuum layer of approximately 20 Å along the $c$-axis to minimize periodic image interactions. Brillouin zone sampling was performed using a Γ-centered $12 \times 12 \times 1$ Monkhorst–Pack $k$-point mesh. A dipole correction was applied along the surface normal to eliminate artificial electrostatic interactions. The surface work function $\Phi$ of $KV_3Sb_5$ was determined using $\Phi = E_{\mathrm{vac}} - E_{\mathrm{F}}$, where $E_{\mathrm{vac}}$ is the planar-averaged electrostatic potential in the vacuum region, and $E_{\mathrm{F}}$ is the Fermi energy[50-52].

## Supporting Information

# High-Performance $KV_3Sb_5/WSe_2$ van der Waals Photodetectors

Yang Yang [1,2], Shaofeng Rao [1,2], Yuxuan Hou [1,2], Jiabo Liu [1,2], Deng Hu [3,4], Yufei Guo [1,2], Jianzhou Zhao [1,2], Hechen Ren [1,2], Zhiwei Wang [3,4 †], and Fan Yang [1,2*]

[1] Center for Joint Quantum Studies and Department of Physics, School of Science, Tianjin University, Tianjin 300350, China

[2] Tianjin Key Laboratory of Low Dimensional Materials Physics and Preparing Technology, Department of Physics, Tianjin University, Tianjin 300350, China

[3] Key Laboratory of Advanced Optoelectronic Quantum Architecture and Measurement, Ministry of Education, School of Physics, Beijing Institute of Technology, Beijing 100081, China

[4] Micronano Center, Beijing Key Lab of Nanophotonics and Ultrafine Optoelectronic Systems, Beijing Institute of Technology, Beijing 100081, China

[†] Corresponding author: zhiweiwang@bit.edu.cn

[*] Corresponding author: fanyangphys@tju.edu.cn

## 1. Structural and Transport Characterization of $KV_3Sb_5$

Figure S1a shows the X-ray diffraction (XRD) pattern of a representative $KV_3Sb_5$ single crystal grown using the self-flux method. Sharp Bragg peaks of the (00$l$) surfaces are clearly observed, demonstrating the high crystallinity and excellent out-of-plane orientation of the crystal.

To characterize the transport properties, a Hall-bar device was fabricated from an exfoliated $KV_3Sb_5$ flake with a thickness of 100 nm, as shown in the inset of Figure S1b. The temperature-dependent resistivity, i.e., $\rho(T)$, exhibits clear metallic behavior over the entire measured temperature range, as shown in Figure S1b. At room temperature, the resistivity is $\rho = 75.6$ μΩ·cm, consistent with previously reported values for high-quality $KV_3Sb_5$ crystals [1]. Upon cooling, the resistivity decreases monotonically and reaches a residual value of $\rho = 1.29$ μΩ cm at $T = 2$ K. The corresponding residual resistance ratio (RRR), defined as defined as $\rho$(300 K)/ $\rho_0$, is therefore calculated to be 58.7, indicating a low level of disorder and impurity scattering in the crystal.

The Figure S1c shows the first derivative of the $\rho(T)$ curve. A kink is observed at $T \approx 75$ K, as marked by the arrow, signaling the onset of the charge-density-wave transition[1-3].

## 2. Magnetoresistance and SdH Oscillation of $KV_3Sb_5$

The semimetallic nature of $KV_3Sb_5$, together with its high carrier mobility, is clearly manifested in its magnetotransport behavior. Figure S2a shows the magnetoresistance (MR) of the $KV_3Sb_5$ Hall-bar device measured at $T = 2$ K. The MR reaches approximately 300% at 14 T and displays pronounced Shubnikov–de Haas (SdH) oscillations superimposed on the classical MR background.

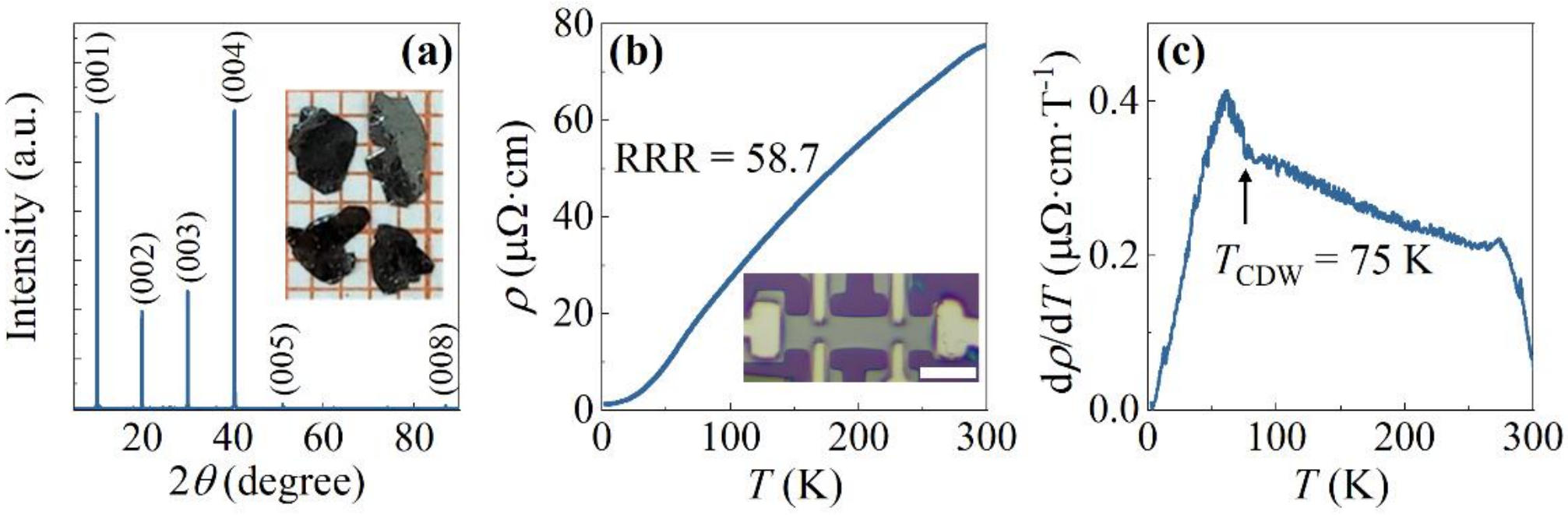

**Figure S1**. a) X-ray diffraction pattern of a typical $KV_3Sb_5$ single crystal. Inset: optical image of as-grown $KV_3Sb_5$ crystals. b) Temperature-dependence resistivity $\rho(T)$ of an exfoliated $KV_3Sb_5$ flake, measured in a Hall-bar geometry, as shown in the inset (scale bar: 10 μm). The residual resistance ratio (RRR) is 58.7, indicative of high crystal quality. c) First derivative of the $\rho(T)$ curve, with the charge-density-wave transition temperature ($T_{\mathrm{CDW}}$) indicated by the arrow.

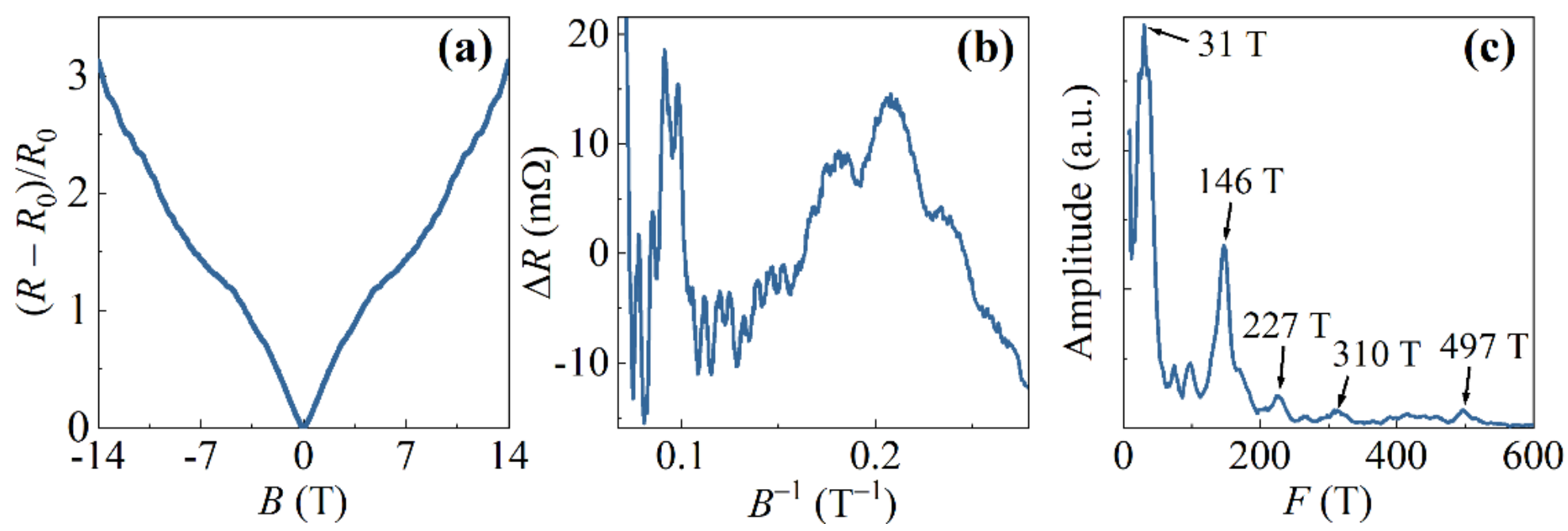

**Figure S2**. a) Magnetoresistance of the $KV_3Sb_5$ Hall-bar device, measured at $T = 2$ K. b) High-field $\Delta R$ data plotted as a function of inverse magnetic field $B^{-1}$, where $\Delta R$ denotes the resistance deviation from the classical magnetoresistance background. Clear Shubnikov–de Haas (SdH) oscillations are observed in the $\Delta R(B^{-1})$ curve. c) Fourier transform of the $\Delta R(B^{-1})$ curve, revealing five distinct SdH frequencies.

To extract the oscillatory component, the high-field classical MR background is modeled using a fifth-order polynomial and subtracted from the raw data. The resulting $\Delta R$ data are plotted as a function of inverse magnetic field $B^{-1}$ in Figure S2b, where clear SdH oscillations are resolved, indicative of a high carrier mobility in $KV_3Sb_5$. Fourier transform analysis of the $\Delta R(B^{-1})$ data reveals five distinct frequencies at 31, 146, 227, 310, and 497 T. The observation of multiple oscillation frequencies indicates the coexistence of several extremal Fermi surface orbits, consistent with the semimetallic nature of $KV_3Sb_5$, where multiple electron- and hole-like pockets contribute to electrical transport[1,4].

## 3. Measurement of Temporal Photocurrent Response

The setup used for measuring the temporal photocurrent response of the $KV_3Sb_5/WSe_2$ device is schematically shown in Figure S3. To reduce the RC time constant of the measurement circuit, the device was connected in series with an external load resistor of 1 kΩ. A 520 nm laser beam, intensity-modulated into rectangular pulses at a repetition frequency of 1 kHz, was incident normal to the device surface. The voltage drop across the load resistor, which is proportional to the photocurrent, was amplified using a low-noise voltage preamplifier (Stanford Research Systems, SR560) with a fixed gain of 100. No hardware filtering was applied during the measurement in order to preserve the intrinsic temporal response of the device. The amplified signal was recorded using a high-bandwidth digital oscilloscope (Keysight 54832B) with a sampling rate of 1 GHz. The rise time ($\tau_r$) and decay time ($\tau_f$) were determined from the rising and falling edges of the transient signals using the 10%–90% criterion.

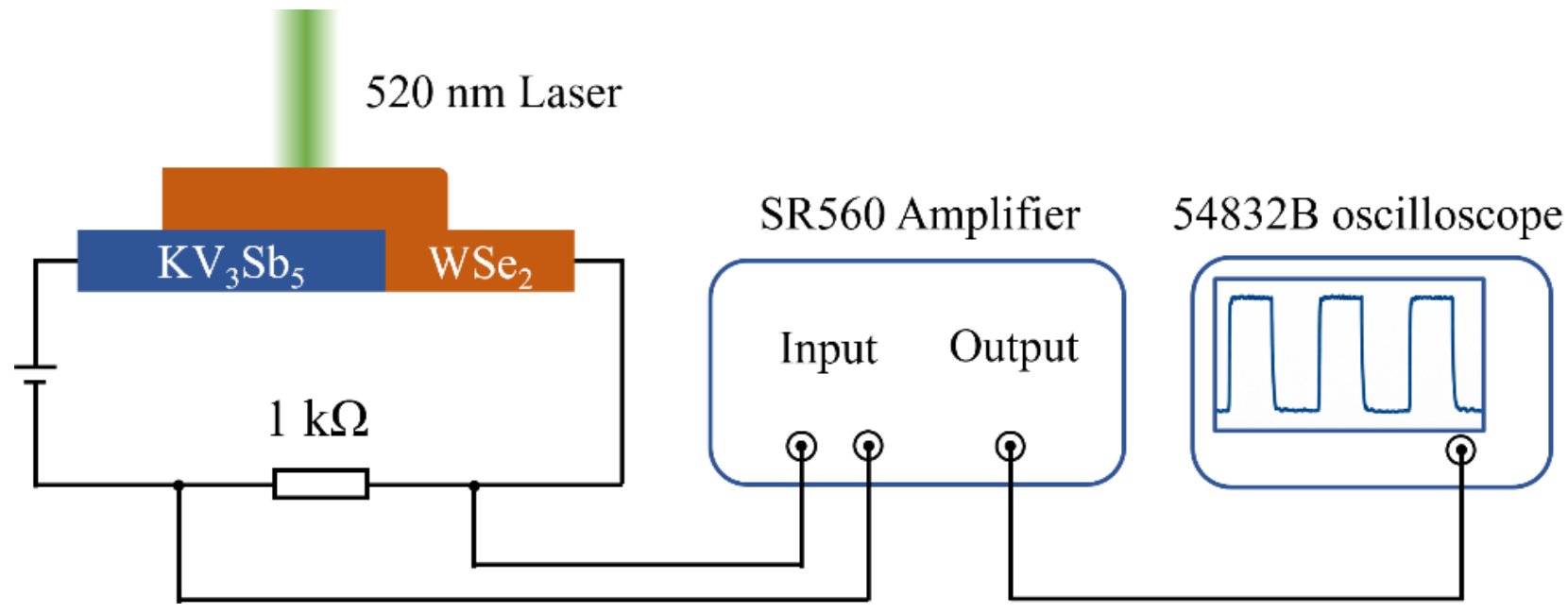

**Figure S3.** Schematic of the experimental setup for measuring the photocurrent response time of the $KV_3Sb_5/WSe_2$ device.

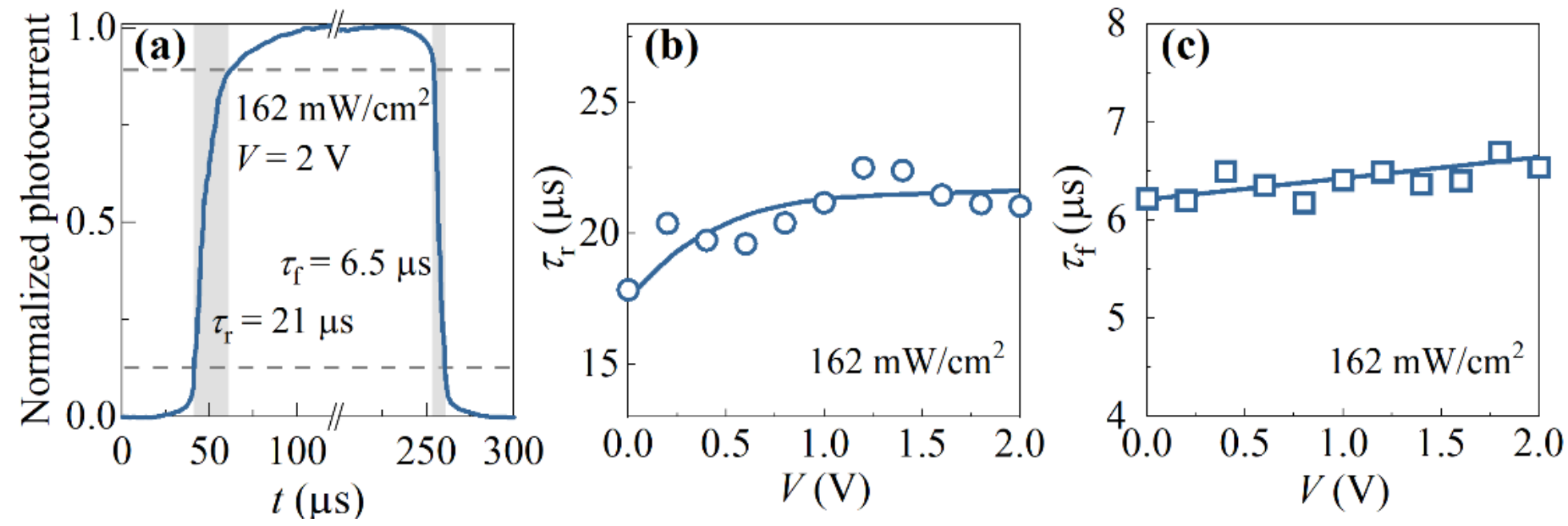

**Figure S4**. a) Temporal photoresponse under 162 mW cm$^{-2}$ illumination at forward bias voltage of 2 V, illustrating the rise and decay times. b) Rise time and c) decay time as a function of bias voltage, measured under 162 mW cm$^{-2}$ illumination.

Temporal photocurrent measurements were performed under different bias voltages. A representative transient response measured under an illumination intensity of 162 mW cm$^{-2}$ at a bias voltage of 2 V is shown in Figure S4a. The extracted $\tau_r$ and $\tau_f$ as functions of bias voltage are summarized in Figures S4b and S4c, respectively. At a bias voltage of 2 V, $\tau_f$ and $\tau_r$ increase by approximately 4% and 18%, respectively, compared to their values measured at zero bias.

## 4. Characterization of an Independent $KV_3Sb_5/WSe_2$ Device

To evaluate the reproducibility of the fabrication process and device performance, a number of $KV_3Sb_5/WSe_2$ devices were fabricated and characterized. While slight variations in performance metrics were observed among different devices, their overall optoelectronic behavior remains highly consistent. As an example, Figure S5 summarizes the results obtained from an independent $KV_3Sb_5/WSe_2$ device (denoted as device S2).

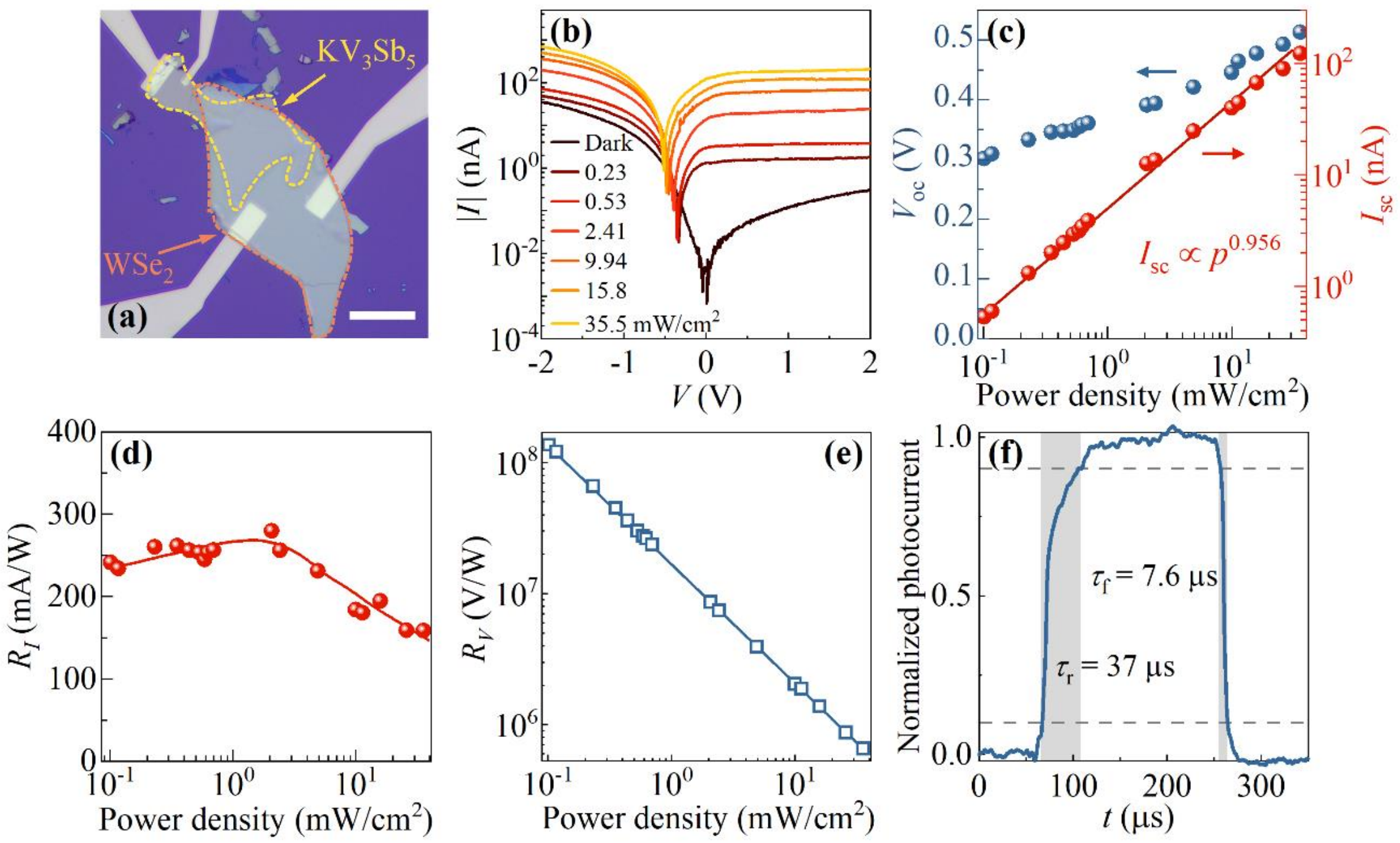

**Figure S5**. a) Optical image of device S2 (scale bar: 40 μm). b) Current–voltage characteristics measured under different illumination intensities. c) Open-circuit voltage ($V_{oc}$) and short-circuit current ($I_{sc}$) as functions of illumination power density. d) Current responsivity and e) voltage responsivity as functions of illumination power density. f) Temporal photoresponse under 56 mW cm$^{-2}$ illumination, with the rise and decay times indicated.

As shown in Figure S5a, device S2 features a larger overlapping region between $KV_3Sb_5$ and $KV_3Sb_5$ compared with the device presented in the main text. In the dark, device S2 exhibits pronounced rectifying $I-V$ characteristics, indicative of Schottky junction formation at the $KV_3Sb_5/WSe_2$ interface. Upon illumination, a substantial increase in conductance is observed over the entire bias voltage range, as shown in Figure S5b.

The short-circuit current of device S2 scales approximately linearly with the illumination intensity, while the open-circuit voltage exhibites a nonlinear behaviour, reaching at about 0.5 V under strong illumination, as summarized in Figure S5c. Correspondingly, the device reaches a maximum current responsivity of approximately 280 mA W$^{-1}$ and a maximum voltage responsivity of about $1.3 \times 10^8$ V W$^{-1}$, as shown in Figures S5d and S5e, respectively.

In addition, device S2 exhibits a fast temporal photoresponse under high illumination intensity, with a rise time of 37 μs and a decay time of 7.6 μs. These results are comparable to those of the primary device discussed in the main text, confirming the reproducibility of both the device fabrication process and the key photodetection characteristics of the $KV_3Sb_5/WSe_2$ heterojunction.